\documentclass[runningheads]{llncs}
\usepackage{graphicx}
\usepackage{multirow}
\begin{document}
\title{Very Fast Keyword Spotting System with Real Time Factor below 0.01}
%
%
\author{Jan Nouza \and
Petr Cerva \and
Jindrich Zdansky}

\authorrunning{J. Nouza et al.}

\institute{Technical University of Liberec, Institute of Information Technologies and Electronics, Studentska 2, Liberec, 46117, Czech Republic \\ \email{\{jan.nouza, petr.cerva, jindrich.zdansky\}@tul.cz}} 
\maketitle              
\begin{abstract}
In the paper we present an architecture of a keyword spotting (KWS) system that is based on modern neural networks, yields good performance on various types of speech data and can run very fast.
We focus mainly on the last aspect and propose optimizations for all the steps required in a KWS design: signal processing and likelihood computation, Viterbi decoding, spot candidate detection and confidence calculation.
We present time and memory efficient modelling by bidirectional feedforward sequential memory networks (an alternative to recurrent nets) either by standard triphones or so called quasi-monophones, and an entirely forward decoding of speech frames (with minimal need for look back).
Several variants of the proposed scheme are evaluated on 3 large Czech datasets (broadcast, internet and telephone, 17 hours in total) and their performance is compared by Detection Error Tradeoff (DET) diagrams and real-time (RT) factors.
We demonstrate that the complete system can run in a single pass with a RT factor close to 0.001 if all optimizations (including a GPU for likelihood computation) are applied.

\end{abstract}

\section{Introduction}

Keyword spotting (KWS) is a frequently used technique in spoken data processing whose goal is to detect selected words or phrases in speech.
It can be applied off-line for fast search in recorded utterances (e.g. telephone calls analysed by police~\cite{p1}), large spoken corpora (like broadcast archives~\cite{p2}), or data collected by call-centres~\cite{p3}.
There are also on-line applications, namely for instant alerting, used in media monitoring~\cite{p4} or in keyword activated mobile services~\cite{p5}. 

The performance of a KWS system is evaluated from two viewpoints.
The primary one is a detection reliability, which aims at missing as few as possible keywords occurring in the audio signal, i.e. to achieve a low miss detection rate (MD), while keeping the number of false alarms (FA) as low as possible.
The second criterion is a speed as most applications require either instant reactions, or they are aimed at huge data (thousands of hours), where it is appreciated if the search takes only a small fraction of their duration.
The latter aspect is often referred to as a real-time (RT) factor and should be significantly smaller than 1.

There are several approaches to solve the KWS task~\cite{p6}.
The simplest and often the fastest one, usually denoted as an \textit{acoustic approach}, utilizes a strategy similar to continuous speech recognition but with a limited vocabulary made of the keywords only.
The sounds corresponding to other speech and noise are modelled and captured by filler units~\cite{p7}.
An \textit{LVCSR approach} requires a very large continuous speech recognition (LVCSR) system that transcribes the audio first and after that searches for the keywords in its text output or in its internal decoder hypotheses arranged in \textit{word lattices}~\cite{p8}.
This strategy takes into account both words from a large representative lexicon as well as inter-word context captured by a language model (LM).
However, it is always slower and fails if the keywords are not in the lexicon and/or in the LM.
A \textit{phoneme lattice approach} operates on a similar principle but with phonemes (usually represented by triphones) as the basic units.
The keywords are searched within the phoneme lattices~\cite{p9}.
The crucial part of all the 3 major approaches consist in assigning a \textit{confidence score} to keyword candidates and setting thresholds for their acceptance or rejection. The basic strategies can be combined to get the best properties of each, as shown e.g. in~\cite{p10,p11}, and in general, they adopt a two-pass scheme.

The introduction of deep neural networks (DNN) into the speech processing domain has resulted in a significant improvement of acoustic models and therefore also in the accuracy of the LVCSR and phoneme based KWS systems.
Various architectures have been proposed and tested, such as feedforward DNNs~\cite{p12}, convolutional (CNN)~\cite{p13} and recurrent ones (RNN)~\cite{p14}.
A combination of the Long Short-Term Memory (LSTM) version of the latter together with the Connectionist Temporal Classification (CTC) method, which is an alternative to the classic hidden Markov model (HMM) approach, have become popular, too.
The CTC provides the location and scoring measure for any arbitrary phone sequence as presented e.g. in~\cite{p15}.
Moreover, modern machine learning strategies, such as training data augmentation or transfer learning have enabled to train KWS also for various signal conditions~\cite{p16} and languages with low data resources~\cite{p17}.

The KWS system presented here is a combination of several aforementioned approaches and techniques.
It allows for searching any arbitrary keyword(s) using an HMM word-and-filler decoder that accepts acoustic models based on various types of DNNs, including feedforward sequential memory networks that are an efficient alternative to RNNs~\cite{p20}.
An audio signal is processed and searched within a single pass in a frame synchronous manner, which means that no intermediate data (such as lattices) need to be precomputed and stored.
This allows for very short processing time (under 0.01 RT) in an off-line mode.
Moreover, the execution time can be further reduced if the same signal is searched repeatedly with a different keyword list.
The system can operate also in an on-line mode, where keyword alerts are produced with a small latency.
In the following text, we will focus mainly on the speed optimization of the algorithms, which is the main and original contribution of this paper.

\section{Brief Description of Presented Keyword Spotting System}

The system models acoustic events in an audio signal by HMMs.
Their smallest units are states.
Phonemes and noises are modelled as 3-state sequences and the keywords as concatenations of the corresponding phoneme models.
All different 3-state models (i.e. physical triphones in a tied-state triphone model) also serve as the fillers.
Hence any audio signal can be modelled either as a sequence of the fillers, or - in presence of any of the keywords – as a sequence of the fillers and the keyword models.
During data processing, the most probable sequences are continuously built by the Viterbi decoder and if they contain keywords, these are located and further managed. 
The complete KWS system is composed of three basic modules.
All run in a frame synchronous manner.
The first one – a \textit{signal processing} module - takes a frame of the signal and computes log-likelihoods for all the HMM states.
The second one – a \textit{state processing} module – controls Viterbi recombinations for all active keywords and filler states.
The third one – a \textit{spot managing} module – focuses on the last states of the keyword/filler models, computes differences in accumulated scores of the keywords and the best filler sequences, evaluates their confidence scores and those with the scores higher than a threshold are further processed.
This scheme assures that the data is processed almost entirely in the forward direction with minimum need for look-back and storage of already processed data.

\section{KWS Speed and Memory Optimizations}
\label{sec:approach}

The presented work extends – in a significant way – the scheme proposed in~\cite{p18}.
Therefore, we will use a similar notation here when explaining optimizations in the three modules.
The core of the system is a Viterbi decoder that handles keywords $w$ and fillers $v$ in the same way, i.e. as generalized units $u$.

\subsection{Signal Processing Module}

It computes likelihoods for each state (senone) using a trained neural network.
This is a standard operation which can be implemented either on a CPU, or on a GPU.
In the latter case, the computation may be more than 1000 times faster.
Yet, we come with another option for a significant reduction in the KWS execution.

The speed of the decoder depends on the number of units that must be processed in each frame.
We cannot change the keyword number but let us see what can be done with the fillers.
Usually, their list is made of all different physical triphones, which means a size of several thousands of items.
If monophones are used instead, the number of fillers would be equal to their number, i.e. it would be smaller by 2 orders and the decoder would run much faster, but obviously with a worse performance. 

We propose an optional alternative solution that takes advantages from both approaches.
We model the words and fillers by something we call quasi-monophones, which can be thought as triphone states mapped to a monophone structure.
In each frame, every quasi-monophone state gets the highest likelihood of the mapped states.
This simple triphone-to-monophone conversion can be easily implemented as an additional layer of the neural network that just takes max values from the mapped nodes in the previous layer.
The benefit is that the decoder handles a much smaller number of different states and namely fillers.
In the experimental section, we demonstrate the impact of this arrangement on KWS system’s speed and performance. 

\subsection{State Processing Module}

The decoder controls a propagation of accumulated scores between adjacent states. At each frame $t$, new score $d$ is computed for each state $s$ of unit $u$ by adding log likelihood $L$ (provided by the previous module) to the higher of the scores in the predecessor states:

\begin{equation}
	d(u,s,t) = L(s,t)+\max_{i=0,1}[d(u,s-i,t-1)]
\end{equation}

Let us denote the score in the unit’s end state $s_E$  as
\begin{equation}
	D(u,t) = d(u,s_E,t)
\end{equation}
and $T(u,t)$ be the frame where this unit’s instance started.
Further, we denote two values $d_{best}$ and $D_{best}$:

\begin{equation}
	d_{best}(t) = \max_{u,s}[d(u,s,t)]
\end{equation} 

\begin{equation}
	D_{best}(t) = \max_{u}[D(u,t)]
\end{equation} 

The former value serves primarily for pruning, the latter is propagated to initial states $s_1$ of all units in the next frame:

\begin{equation}
	d(u,s_{1},t+1) = L(s_1,t+1)+\max[D_{best}(t),d(u,s_{1},t)]
\end{equation}

\subsection{Spot Managing Module}

This module computes acoustic scores $S$ for all words $w$ that reached their last states.
This is done by subtracting these two accumulated scores:

\begin{equation}
\label{eq6}
	S(w,t) = D(w,t) - D_{best}(T(w,t)-1)
\end{equation}

The word score $S(w,t)$ needs to be compared with score $S(v_{string},t)$ that would be achieved by the best filler string $v_{string}$ starting in frame $T(w,t)$ and ending in frame $t$. 

\begin{equation}
\label{eq7}
	R(w,t) = S(v_{string},t) - S(w,t)
\end{equation}

In~\cite{p18}, the first term in eq.~\ref{eq7} is computed by applying the Viterbi algorithm within the given frame span to the fillers only.
Here, we propose to approximate its value by this simple difference: 

\begin{equation}
\label{eq8}
	S(v_{string},t) \cong  D_{best}(t) - D_{best}(T(w,t)-1)
\end{equation}

The left side of eq.~\ref{eq8} equals exactly the right one if  the Viterbi backtracking path passes through frame $T(w,t)$, which can be quickly checked.
A large experimental evaluation showed that this happens in more than 90 \% cases.
In the remaining ones, the difference is so small that it has a negligible impact on further steps.

Hence, by substituting from eq.~\ref{eq6} and eq.~\ref{eq8} into eq.~\ref{eq7} we get:

\begin{equation}
\label{eq9}
	R(w,t) = D_{best}(t) - D(w,t)
\end{equation}

The value of $R(w,t)$ is related to the confidence of word $w$ being detected in the given frame span.
We just need to normalize it and convert it to a human-understandable scale where number 100 means the highest possible confidence.
We do it in the following way:

\begin{equation}
\label{eq10}
	C(w,t) = 100 - k\frac{R(w,t)}{(t-T(w,t))N_S(w)}
\end{equation}

The $R$ value is divided by the word duration (in frames) and its number of HMM states $N_s$, which is further multiplied by constant $k$ before subtracting the term from 100.
The constant influences the range of the confidence values.
We set it so that the values are easily interpretable by KWS system users (see section~\ref{sec:evalres}).

The previous analysis shows that the spot managing module can be made very simple and fast.
In each frame, it just computes eq.~\ref{eq9} and~\ref{eq10} and the candidates with the confidence scores higher than a set threshold are registered in a time-sliding buffer (10 to 20 frames long).
A simple filter running over the buffer content detects the keyword instance with the highest score and sends it to the output. 

\subsection{Optimized Repeated Run}

In many practical applications, the same audio data is searched repeatedly, usually with different keyword lists (e.g. during police investigations).
In this case, the KWS system can run significantly faster if we store all likelihoods and two additional values ($d_{best}$ and $D_{best}$) per frame.
In the repeated run, the signal processing part is skipped over and the decoder can process only the keywords because all information needed for optimal pruning and confidence calculation is covered by the 2 above mentioned values. 

\section{System and Data for Evaluation}

\subsection{KWS System}

The KWS system used in the experiments is written in C language and runs on a PC (Intel Core i7-9700K).
In some tasks we employ also a GPU (GeForce RTX 2070 SUPER) for likelihood computation. 

We tested 2 types of acoustic models (AM) based on neural networks.
Both accept 16 kHz audio signals, segmented into 25ms long frames and preprocessed to 40 filter bank coefficients.
The first uses a 5-layer feedforward DNN trained on some 1000 hours of Czech data (a mix of read and broadcast speech).
The second AM utilizes a bidirectional feedforward sequential memory network (BFSMN) similar to that described in~\cite{p20}.
We have been using it as an effective alternative of RNNs.
In our case, it has 11 layers, each covering 4 left and 4 right temporal contexts.
This AM was trained on the same source data augmented by about 400 hours of (originally) clean speech that passed through different codecs~\cite{p21}.
For both types of the NNs we have trained triphone AMs, for the second also a monophone and quasi-monophone version.

\subsection{Dataset for Evaluation}
\label{sec:dataset}
 
Three large datasets have been prepared for the evaluation experiments, each covering a different type of speech (see also Table~\ref{tab:dataset}).
The Interview dataset contains 10 complete Czech TV shows with two-persons talking in a studio.
The Stream dataset is made of 30 shows from Internet TV Stream.
We selected the shows with heavy background noise, e.g. Hudebni Masakry (Music Masacres in English).
The Call dataset covers 53 telephone communications with call-centers (in separated channels) and it is a mix of spontaneous (client) and mainly controlled (operator) speech.
All recordings have been carefully annotated with time information (10 ms resolution) added to each word.

\begin{table}[ht]
\centering
\caption{Datasets for evaluation and their main parameters.}\label{tab:dataset}
\begin{tabular}{|l|c|c|c|c|}
\hline
\bfseries Dataset & \bfseries Speech type & \bfseries Signal type & \bfseries Total duration [min] & \bfseries \# keywords\\
\hline
\hline
Interview & planned & studio &	272 & 3524 \\
\hline
Stream &  informal & heavy noise	& 157 & 1454 \\
\hline
Call &  often spontaneous & telephone &	613 & 2935 \\
\hline
\end{tabular}
\end{table}

\section{Experimental Evaluation}
\label{sec:eval}

\subsection{Keyword List}

Our goal was to test the system under realistic conditions and, at the same time, to get statistically conclusive results.
A keyword list of 156 word lemmas with 555 derived forms was prepared for the experiments.
For example, in case of keyword ''David'' we included its derived forms ''David'', ''Davida'', ''Davidem'', ''Davidovi'', etc. in order to avoid false alarms caused by words being substrings of others.
The list was made by combining 80 most frequent words that occurred in each of the datasets, from which some were common and some appeared only in one set.
The searched word forms had to be at least 4 phonemes long.
The mean length of the listed word forms was 6.9 phonemes.
The phonetic transcriptions were automatically extracted from a 500k-word lexicon used in our LVCSR system.

\subsection{Filler Lists}

The list of fillers was created automatically for each acoustic model.
The triphone DNN model generated 9210 fillers and the triphone BFSMN produced 10455 of them.
In contrast to these large numbers, the monophone and quasi-monophone BFSMN model had only 48 fillers (representing 40 phonemes + 8 noises).

\subsection{Evaluation conditions and metrics}

A word was considered  correctly detected if the spotted word-form belonged to the same lemma as the word occurring in the transcription at the same instant - with tolerance ±0.5 s.
Otherwise it was counted as a false alarm.
For each experiment we computed Missed Detection (MD) and False Alarm (FA) rates as a function of acceptance threshold value, and drawn a Detection Error Tradeoff (DET) diagram with a marked Equal Error Rate (EER) point position. 

\subsection{Evaluation results}
\label{sec:evalres}

\begin{figure}[ht!]
\centering
\includegraphics[scale=0.52]{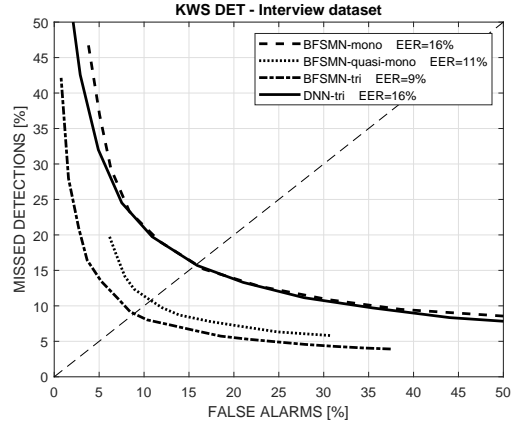}
\caption{KWS results for the Interview dataset in form of DET curves drawn for 4 investigated neural network structures.} \label{fig1}
\end{figure}

The Interview dataset was used as a development data, on which we experimented with various models, system arrangements and also user preferences.
In accord with them, the internal constant $k$ occurring in eq.~\ref{eq10} was set to locate the confidence score equal to 75 close to the EER point. 
The first part of the experiments focused on the accuracy of the created acoustic models.
We tested the triphone DNN and 3 versions of the BFSMN one.
Their performance is illustrated by DET curves in Fig.~\ref{fig1}, where also the EER values are displayed.
It is evident that the BFSMN-tri model performs significantly better than the DNN one, which is mainly due to its wider context span.
This is also a reason why even its monophone version has performance comparable to the DNN-tri one.
The proposed quasi-monophone BFSMN model shows the second best performance but the gap between it and the best one is not that crucial, especially if we take into account its additional benefits that will be discussed later.

\begin{figure}[ht!]
\centering
\includegraphics[scale=0.52]{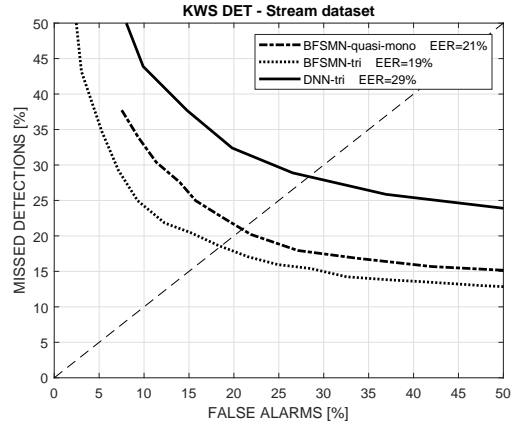}
\caption{DET curves compared for 3 models on the Stream dataset} \label{fig2}
\end{figure}

\begin{figure}[ht!]
\centering
\includegraphics[scale=0.52]{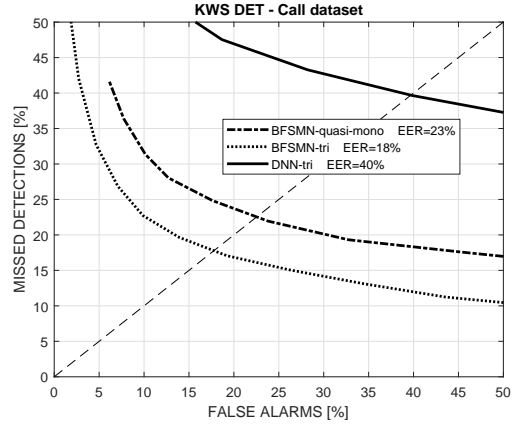}
\caption{DET curves compared for 3 models on the Call dataset} \label{fig3}
\end{figure}

Similar trends can be seen also in Fig.~\ref{fig2} and Fig.~\ref{fig3} where we compare the same models (excl. the monophone BFSMN) on the Stream and Call datasets.
In both cases, the performance of all the models was worse (when compared to that of the Interview set) as it can be seen from the positions of the curves and the EER values. This is due to the character of speech and signal quality as explained is section~\ref{sec:dataset}.
Yet, we can notice the positive effect of the training of the BFSMN models on the augmented data (with various codecs), especially on the Call dataset.
Again, the gap between the best triphone and the proposed quasi-monophone version seems to be not that critical.

Now, we shall focus on the execution time of the proposed scheme.
As explained in section~\ref{sec:approach}, the three modules of the KWS system can be split into 2 parts: the first with the signal processing module, the second with the remaining two.
Both can run together on a PC (in a single thread), or if extremely fast execution is required, the former can be implemented on a GPU.
We tested both approaches and measured their RT factors.
Similar measurements (across all the tree datasets) were performed also in the second part for all the proposed variants and operation modes (see Table~\ref{tab:times} for results.)
The total RT factor is obtained by adding the values for selected options in each of the two parts.

\begin{table}[ht!]
\centering
\caption{Execution times for proposed KWS variants expressed as RT factors.}\label{tab:times}
\begin{tabular}{|l|c|}
\hline
\bfseries System part, variant, mode & \bfseries Real-Time factor\\
\hline
\hline
\multicolumn{2}{|c|}{Part 1 (signal proc. module)} \\
\hline
on CPU & 0.12 \\
\hline
on GPU & 0.0005 \\
\hline
\multicolumn{2}{|c|}{Part 2 (rest of KWS system)} \\
\hline
triphone BFSMN & 0.012 \\
\hline
quasi-mono BFSMN & 0.002 \\
\hline
triphone BFSMN, repeated & 0.009 \\
\hline
quasi-mono BFSMN, repeated & 0.001 \\
\hline
\end{tabular}
\end{table}

Let us remind that the proposed quasi-monophone model performs slightly worse but it offers two practical benefits: a) a speed that can get close to 0.001 RT (if a GPU is used for likelihood computation) and b) a small disk memory consumption in case of repeated runs (with different keywords) because only 48x3+2=146 float numbers per frame need to be stored.
Moreover, the speed of the proposed KWS system is only slightly influenced by the number of keywords.
A test made with 10.000 keywords (instead of 555 ones used in the main experiments) showed only twice slower performance.

\section{Conclusion}

In this contribution we focus mainly on the speed aspect of a modern KWS system, but at the same time we aim at the best performance that is available thanks to the advances in deep neural networks.
The used BFSMN architecture has several benefits for practical usage.
In contrast to more popular RNNs, it can be efficiently and fast trained on a large amount (several thousands of hours) of audio and at the same time yields performance comparable to more complex RNNs and LSTMs as shown in~\cite{p20}.
Its phoneme accuracy is high (due its large internal context) so that it fits both to acoustic KWS systems as well as to standard speech-to-text LVCSR systems.
The latter means that it is well suited for a tandem KWS scheme where a user requires that the sections with detected keywords are immediately transcribed by a LVCSR system.
In our arrangement this can be done very effectively by reusing some of the precomputed data.
(Let us recall that if we use the quasi-monophones, their values are just max values from the original triphone neural network and hence both acoustic models can be implemented by the same network with an additional layer.)

The results presented in section~\ref{sec:eval} allow for designing an optimal configuration that takes into account the three main factors: accuracy, speed and cost.
If the main priority is accuracy and not the speed, the KWS system can run on a standard PC and process data with a RT factor about 0.1.
When very large amounts of records must be processed within very short time then the addition of a GPU and the adoption of the proposed quasi-monophone approach will allow for completing the job in time that can be up to 3 orders shorter than the audio duration. 

We evaluated the performance on Czech datasets as these were available with precise human checked transcriptions.
Obviously, the proposed architecture is language independent and we plan to utilize it for other languages investigated in our project.

\subsubsection*{Acknowledgments.}
This work was supported by the Technology Agency of the Czech Republic (Project No. TH03010018).

\end{document}